\pdfoutput=1
\documentclass[twocolumn,prl,superscriptaddress,longbibliography]{revtex4-1}
\usepackage{graphicx,amsmath,amssymb}
\usepackage[colorlinks=true, citecolor=blue, urlcolor=blue, linkcolor=red]{hyperref}
\renewcommand{\section}[1]{{\par\it #1.---}\ignorespaces}

\begin{document}
\title{Low-temperature Quantum Metrology Enhanced by Strong Couplings}
\author{Ze-Zhou Zhang}
\affiliation{Key Laboratory of Quantum Theory and Applications of Ministry of Education, Lanzhou Center for Theoretical Physics and Key Laboratory of Theoretical Physics of Gansu Province,
Lanzhou University, Lanzhou 730000, China}
\author{Hong-Gang Luo}
\affiliation{Key Laboratory of Quantum Theory and Applications of Ministry of Education, Lanzhou Center for Theoretical Physics and Key Laboratory of Theoretical Physics of Gansu Province,
Lanzhou University, Lanzhou 730000, China}
\author{Wei Wu}
\email{wuw@lzu.edu.cn}
\affiliation{Key Laboratory of Quantum Theory and Applications of Ministry of Education, Lanzhou Center for Theoretical Physics and Key Laboratory of Theoretical Physics of Gansu Province,
Lanzhou University, Lanzhou 730000, China}

\begin{abstract}
Equilibrium probes have been widely used in various noisy quantum metrology schemes. However, such an equilibrium-probe-based metrology scenario severely suffers from the low-temperature-error-divergence problem in the weak-coupling regime. To circumvent this limit, we propose a strategy to eliminate the error-divergence problem by utilizing the strong coupling effects, which can be captured by the reaction-coordinate mapping. The strong couplings induce a noncanonical equilibrium state and greatly enhance the metrology performance. It is found that our metrology precision behaves as a polynomial-type scaling relation, which suggests the reduction of temperature can be used as a resource to improve the metrology performance. Our result is sharply contrary to that of the weak-coupling case, in which the metrology precision exponentially decays as the temperature decreases. Paving a way to realize a high-precision noisy quantum metrology at low temperatures, our result reveals the importance of the non-Markovianity in quantum technologies.
\end{abstract}
\maketitle

\section{Introduction}\label{sec:sec1}
Quantum metrology is a rapidly developing research field, which aims at surpassing the precision limit set by classical statistics~\cite{RevModPhys.90.035005,RevModPhys.89.035002,RevModPhys.92.015004}. It has been widely demonstrated that some quantum resources, such as quantum entanglement~\cite{Lachance-Quirion425,PhysRevLett.122.123605,Zou6381} and quantum squeezing~\cite{PhysRevD.23.1693,PhysRevLett.118.140401,PhysRevLett.119.193601}, are able to boost the metrology performance for the noiseless ideal case. However, in any practical metrology scenarios, the probe, in which the parameter of interest is intrinsic or externally imprinted, inevitably interacts with its surrounding bath~\cite{https://doi.org/10.1002/qute.202300218,PhysRevLett.79.3865,PhysRevLett.120.140501,PhysRevLett.109.233601,PhysRevLett.133.090801}. Thus, the effect of decoherence should be taken into account. Roughly speaking, there are two kinds of quantum probe used in the studies of noisy quantum metrology: one is the equilibrium probe~\cite{Paris_2016,PhysRevLett.114.220405,PhysRevA.108.032220,Potts2019fundamentallimits,PhysRevLett.133.040802} and the other one is the nonequilibrium dynamical probe~\cite{Haase_2018,PhysRevA.103.L010601,PhysRevApplied.15.054042,Tamascelli_2020}. If the decoherence time is shorter than the probe-bath interaction time, the probe is fully thermalized. In this case, the signal-to-noise ratio (SNR) of the estimated parameter is obtained from the equilibrium state of the probe. If the interaction time is very short, the coherence can be partially reserved, and the SNR is then read out from dynamical signals.

Compared with the nonequilibrium dynamical probe, the equilibrium one has certain advantages. For example, the equilibrium-probe scheme does not require any precise dynamical control to find the optimal interrogation time~\cite{PhysRevA.95.012109,PhysRevA.97.012125,PhysRevA.97.012126,PhysRevA.100.032318} and is commonly universal to different initial states. However, except for very a few solvable models, an analytical expression for the probe's equilibrium state is generally difficult to obtained. The utilization of numerical tools may be a possible solution~\cite{10.1063/5.0011599,10.1063/5.0140002,VELIZHANIN2008325}, but it relies heavily on computing resources. The other way to overcome this restriction is the employment of the weak-coupling approximation, which assumes the coupling between the probe and the bath is so weak that the probe experiences a canonical thermalization. Under this approximate framework, the equilibrium state of the probe can be described by a canonical equilibrium (Gibbs) state, which greatly simplifies the difficulties in calculating the SNR. Unfortunately, the weak-coupling approximation neglects the system-bath correlations, which is commonly viewed as a non-Markovian effect~\cite{PhysRevLett.104.250401,doi:10.1143/JPSJ.81.063301,PhysRevA.90.032114}. This treatment gives rise to a divergent metrology error at low temperature, which has been pointed out in various quantum thermometries~\cite{Paris_2016,Potts2019fundamentallimits,PhysRevA.96.062103,PhysRevA.109.042417,PhysRevResearch.2.033394,PhysRevLett.128.040502} and the quantum metrology for Hamiltonian parameters~\cite{Gabbrielli2018,PhysRevA.90.022111,REN2022105542,PhysRevA.94.042121,PhysRevLett.133.040802}. It is highly desirable to develop an alternative scheme to eliminate this problem.

In this Letter, we propose a strategy to improve the precision of an equilibrium-probe-based noisy frequency estimation scenario via strong couplings. By using the reaction-coordinate mapping~\cite{LIANG2007296,PhysRevA.90.032114,10.1063/1.4940218,10.1063/1.3532408}, in the strong-coupling regime, we find the noncanonical characteristics occurs in the equilibrium state of the probe that are composed of $N$ spin-$1/2$ particles. Such a noncanonicality is induced by the non-Markovianity~\cite{PhysRevA.90.032114,PhysRevA.104.052617,Strasberg_2016,doi:10.1143/JPSJ.81.063301,10.1063/1.4940218,PhysRevResearch.4.023169,PhysRevE.90.022122}, which is neglected in the usual weak-coupling treatment. At low temperatures, it is found that the SNR from our strategy can maintain as a temperature-independent constant or behave as a polynomial scaling relation according to different spin numbers. This result is sharply contrary to that of the weak-coupling case, in which the SNR exponentially decays when the temperature is lowered, and completely remove the error-divergence problem.

\section{Weak-coupling approximation}
In this Letter, we consider a quantum metrology scenario of noisy frequency estimation, which is experimentally related to the atomic spectroscopy~\cite{Haase_2018,Wang_2017,PhysRevX.14.011033}. In our task, we estimate the frequency of a quantum probe, which is composed of multiple spin-$1/2$ particles with the Hamiltonian $\hat{H}_{\text{s}}=\epsilon \hat{J}_{z}$. Here, $\epsilon$ is the parameter to be determined and $\hat{J}_{\upsilon}\equiv\frac{1}{2}\sum_{n=1}^{N}\hat{\sigma}_{n}^{\upsilon}$ with $\upsilon=x,y,z$ are the collective spin operators. Such an estimation of $\epsilon$ can be realized via a unitary dynamics in the ideal noiseless case~\cite{PhysRevLett.79.3865}. However, such a unitary metrology scheme breaks down when the inevitable noises from the surrounding environment are taken into account. In the noisy metrological case, we assume the influences of environmental noises can be described via a linear interaction between the probe and a thermal bosonic bath $\hat{H}_{\text{b}}=\sum_{k}\omega_{k}\hat{b}_{k}^{\dagger}\hat{b}_{k}$, which is in a thermal equilibrium at the temperature $T=1/\beta$ ($k_{\text{B}}=1$). The total Hamiltonian reads ($\hbar=1$)
\begin{equation}\label{eq:eq1}
\hat{H}=\hat{H}_{\text{s}}+\hat{H}_{\text{b}}+\hat{J}_{x}\sum_{k}g_{k}(\hat{b}_{k}^{\dagger}+\hat{b}_{k}),
\end{equation}
where $\hat{b}_{k}$ and $\hat{b}_{k}^{\dagger}$ are the annihilation and creation operators of the $k$th bosonic mode with frequency $\omega_{k}$, respectively. Parameters $g_{k}$ denote the coupling strength between the probe and the $k$th bosonic mode. The frequency dependence of the interaction strengths is encoded into the spectral density $J(\varpi)\equiv\sum_{k}g_{k}^{2}\delta(\varpi-\omega_{k})$. The explicit expression of $J(\varpi)$ will not be addressed here, because our result is universal to the form of $J(\varpi)$~\cite{10.1063/1.3532408}.

Assuming the probe-bath interaction time is sufficiently long, the probe shall evolve to its long-time steady state. This equilibrium state is approximated as a Gibbs state $\rho_{\text{s}}^{\text{weak}}(\infty)=e^{-\beta \hat{H}_{\text{s}}}/Z$ with $Z\equiv\text{Tr}(e^{-\beta \hat{H}_{\text{s}}})$, if one takes the weak-coupling approximation~\cite{PhysRevE.90.022122,Breuer,PhysRevLett.127.250601,10.1063/1.4722336}. Under this approximation, by measuring the observable $\hat{J}_{z}$, one can easily find the SNR $\mathbb{S}=|\partial_{\epsilon}\langle \hat{J}_{z}\rangle|^{2}/(\langle \hat{J}_{z}^{2}\rangle-\langle \hat{J}_{z}\rangle^{2})$ is given by
\begin{equation}\label{eq:eq2}
\mathbb{S}_{\text{weak}}=\frac{N\beta^{2}}{2+2\cosh(\beta\epsilon)}.
\end{equation}
From the above expression, one proves that
\begin{equation}\label{eq:eq3}
\lim_{T\rightarrow 0}\mathbb{S}_{\text{weak}}\propto e^{-\epsilon/T},
\end{equation}
which means the SNR of the equilibrium-state probe exponentially decay as $\epsilon/T\rightarrow 0$. This is the outstanding error-divergence problem~\cite{Paris_2016,Potts2019fundamentallimits,PhysRevA.96.062103,PhysRevA.109.042417,PhysRevResearch.2.033394,PhysRevLett.128.040502,Gabbrielli2018,PhysRevA.90.022111,REN2022105542,PhysRevA.94.042121,PhysRevLett.133.040802}, which severely limits the equilibrium-probe-based quantum metrology at low temperature.

\begin{figure}
\centering
\includegraphics[angle=0,width=0.45\textwidth]{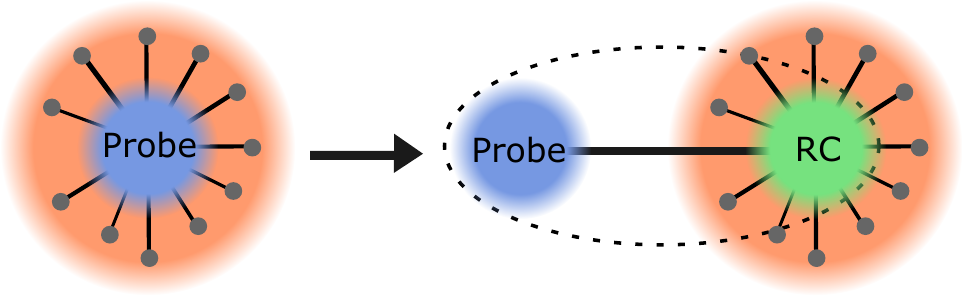}
\caption{Diagrammatic sketch of the reaction-coordinate mapping. In the original Hamiltonian, a quantum probe with Hamiltonian $\hat{H}_{\text{s}}$, which is large spin system, is directly coupled to a thermal bosonic bath. By applying the reaction-coordinate mapping, the probe interacts with a RC mode within an extend composite system with Hamiltonian $\hat{\mathcal{H}}_{\text{s}}$, which is in turn weakly connected to a residual thermal bath. In the reaction-coordinate-mapping picture, the equilibrium of the probe is corrected as a noncanonical state $\rho_{\text{s}}(\infty)\propto\text{Tr}_{\text{RC}}[\exp(-\beta \hat{\mathcal{H}}_{\text{s}})]$, compared with the weak-coupling result $\rho_{\text{s}}^{\text{weak}}(\infty)\propto \exp(-\beta \hat{H}_{\text{s}})$.}\label{fig:fig1}.
\end{figure}

To overcome the above problem, one needs to go beyond the limitation of the weak-coupling approximation, because a strong probe-bath coupling is able to induce the non-Markovian and the bound-state effects, which are beneficial to improve the metrology performance~\cite{PhysRevA.102.032607,PhysRevLett.109.233601,PhysRevA.88.035806,PhysRevA.103.L010601,PhysRevApplied.15.054042}. However, the exactly analytical expression for the equilibrium state of the probe is general difficult to obtain. Fortunately, as demonstrated in many previous studies~\cite{PhysRevA.90.032114,10.1063/1.4940218,10.1063/5.0207028,PRXQuantum.4.020307,10.1063/1.4866769}, the long-time behaviour of a quantum dissipative dynamics can be faithfully captured by the reaction-coordinate-mapping approach, which provides a simple but accurate expression to describe the probe's equilibrium state when suitable parameters in the spectral density are chosen. Next, we apply this technique to our metrology scenario.

\section{Reaction-coordinate-mapping approach}
The reaction coordinate transformation, which is an orthogonal transformation, maps the original Hamiltonian $\hat{H}$ to a dynamically equivalent Hamiltonian $\hat{\mathcal{H}}$~\cite{SupplementalMaterial}. As displaced in Fig.~\ref{fig:fig1}, in the picture of the reaction-coordinate mapping, the probe is directly interacts with one collective reaction coordinate (RC) mode, which is in turn coupled to a residual environment. The mapped Hamiltonian reads $\hat{\mathcal{H}}=\hat{\mathcal{H}}_{\text{s}}+\hat{\mathcal{H}}_{\text{b}}+\hat{\mathcal{H}}_{\text{int}}+\hat{\mathcal{H}}_{\text{c}}$, where
\begin{equation}\label{eq:eq4}
\hat{\mathcal{H}}_{\text{s}}=\hat{H}_{\text{s}}+\omega \hat{a}^{\dagger}\hat{a}+g\hat{J}_{x}(\hat{a}^{\dagger}+\hat{a}).
\end{equation}
Here, the coupling strength $g$ and the frequency of the RC mode $\omega$ are given in terms of $J(\varpi)$~\cite{PhysRevA.108.032220,10.1063/5.0207028}. The residual environment reads $\hat{\mathcal{H}}_{\text{b}}=\sum_{k}\tilde{\omega}_{k}\hat{a}_{k}^{\dagger}\hat{a}_{k}$, $\hat{\mathcal{H}}_{\text{int}}=(\hat{a}^{\dagger}+\hat{a})\sum_{k}\tilde{g}_{k}(\hat{a}_{k}^{\dagger}+\hat{a}_{k})$ is the interaction Hamiltonian and $\hat{\mathcal{H}}_{\text{c}}=(\hat{a}^{\dagger}+\hat{a})^{2}\sum_{k}\tilde{g}_{k}^{2}/\tilde{\omega}_{k}$ is the counterterm.

Through this approach, the probe $\hat{H}_{\text{s}}$ is incorporated within an extend composite system $\hat{\mathcal{H}}_{\text{s}}$. One can realize an arbitrarily strong probe-bath coupling in the original Hamiltonian $\hat{H}$, while having arbitrarily small coupling between $\mathcal{\hat{H}}_{\text{s}}$ and $\mathcal{\hat{H}}_{\text{b}}$ in the mapped one~\cite{SupplementalMaterial}. Under this condition, one finds the long-time steady state of the composite system is well-described by $\varrho_{\text{s}}(\infty)=e^{-\beta \hat{\mathcal{H}}_{\text{s}}}/\mathcal{Z}$ with $\mathcal{Z}=\text{Tr}(e^{-\beta \hat{\mathcal{H}}_{\text{s}}})$. After tracing out the RC degrees of freedom, the steady state for the probe can be obtained accordingly as $\rho_{\text{s}}(\infty)=\text{Tr}_{\text{RC}}[\varrho_{\text{s}}(\infty)]$. The validity for the noncanonical state of $\rho_{\text{s}}(\infty)$ has been demonstrated via some numerically rigorous methods in Refs.~\cite{PhysRevA.90.032114,10.1063/1.4940218,10.1063/5.0207028,PRXQuantum.4.020307}. Compared with the weak-coupling case, the phenomenon of the noncanonical equilibration is captured by the reaction-coordinate-mapping treatment, which faithfully reflects the probe-bath correlations or the non-Markovian effects in strong-coupling regimes.

With the expression of $\rho_{\text{s}}(\infty)$ at hand, the SNR can be easily obtained via numerically diagonalizing $\hat{\mathcal{H}}_{\text{s}}$. To make a comparison and to build a clear physical picture, we also provide an analytical result from the generalized rotating-wave approximation (GRWA)~\cite{PhysRevLett.99.173601,PhysRevA.86.015803,PhysRevA.91.013814,PhysRevA.94.012317}. Good agreement is found (see the comparisons in Supplemental Materials~\cite{SupplementalMaterial}).

\begin{figure}
\centering
\includegraphics[angle=0,width=0.485\textwidth]{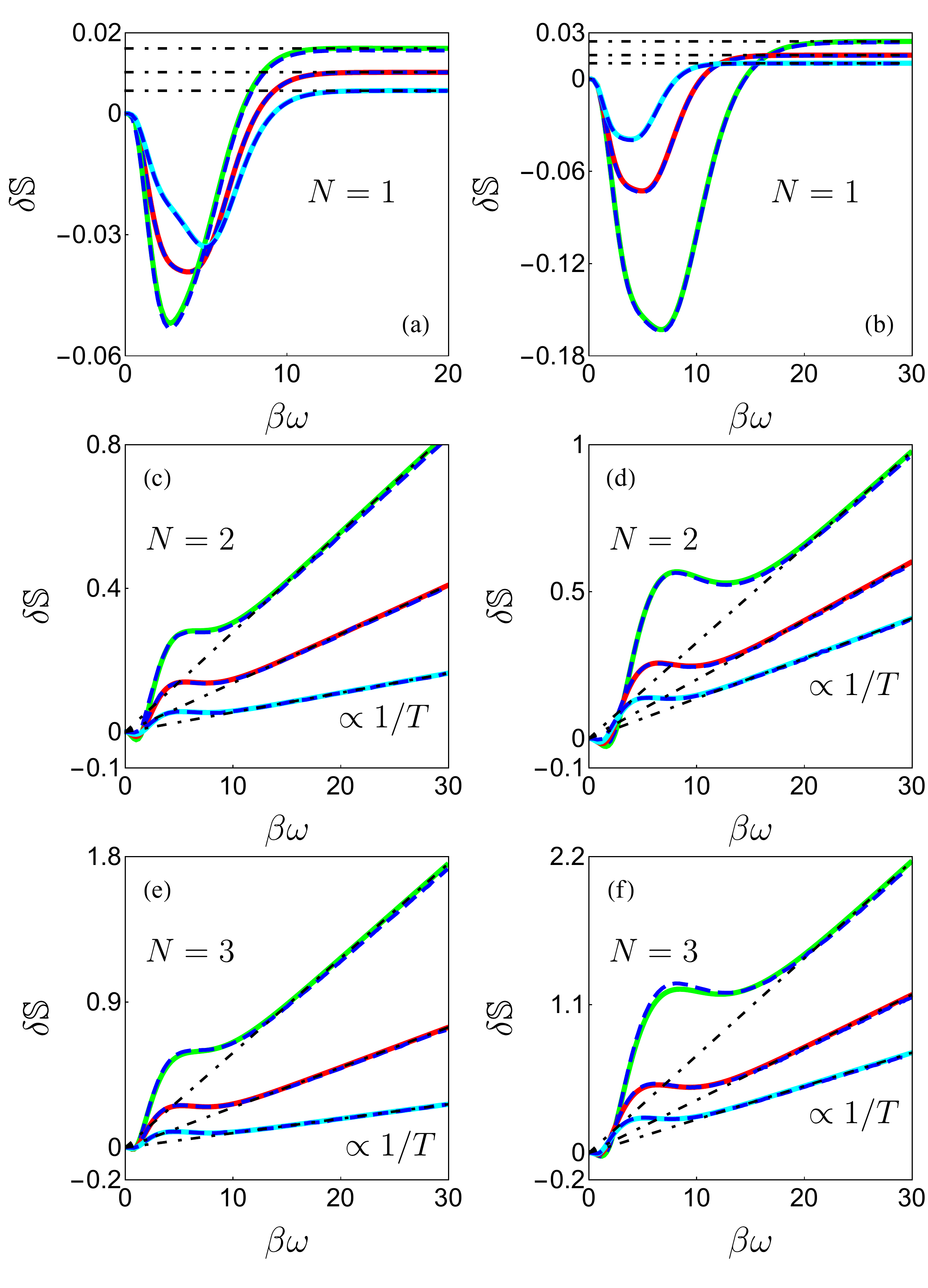}
\caption{Left panel: $\delta \mathbb{S}$ versus $\beta\omega$ with $\epsilon=\omega$ for different parameters $(g/\omega, N)=(0.3,1)$ (cyan line), $(g/\omega, N)=(0.4,1)$ (red line) and $(g/\omega, N)=(0.5,1)$ (green line) in (a); $(g/\omega, N)=(0.2,2)$ (cyan line), $(g/\omega, N)=(0.3,2)$ (red line) and $(g/\omega, N)=(0.4,2)$ (green line) in (c), and $(g/\omega, N)=(0.2,3)$ (cyan line), $(g/\omega, N)=(0.3,3)$ (red line) and $(g/\omega, N)=(0.4,3)$ (green line) in (e). Right panel: $\delta \mathbb{S}$ versus $\beta\omega$ for $\epsilon/\omega=1$ (cyan line), $\epsilon/\omega=0.8$ (red line) and $\epsilon/\omega=0.6$ (green line) with $g/\omega=0.4$ and $N=1$ in (b); $g/\omega=0.3$ and $N=2$ in (d); $g/\omega=0.3$ and $N=3$ in (f). The blue dashed lines are analytical results predicted by the GRWA method, while the black dot-dashed lines are approximate results from Eq.~(\ref{eq:eq7}) and Eq.~(\ref{eq:eq8}) with $E_{g}$ from the GRWA approach.}\label{fig:fig2}
\end{figure}

\section{Generalized rotating-wave approximation}
In the GRWA treatment, we first apply an unitary transformation to $\hat{\mathcal{H}}_{\text{s}}$ as $\hat{\mathcal{H}}'_{\text{s}}=e^{\lambda \hat{J}_{x}(\hat{a}^{\dagger}-\hat{a})}\hat{\mathcal{H}}_{\text{s}}e^{-\lambda \hat{J}_{x}(\hat{a}^{\dagger}-\hat{a})}$ with $\lambda$ being self-consistently determined by minimizing the ground-state energy. The transformed Hamiltonian is
\begin{equation}\label{eq:eq5}
\begin{split}
\hat{\mathcal{H}}'_{\text{s}}=&(\omega\lambda^{2}-2g\lambda)\hat{J}_{x}^{2}+\omega \hat{a}^{\dagger}\hat{a}+(g-\omega\lambda)J_{x}(\hat{a}^{\dagger}+\hat{a})\\
&+\epsilon\hat{J}_{z}\cosh[\lambda(\hat{a}^{\dagger}-\hat{a})]-i\hat{J}_{y}\sinh[\lambda(\hat{a}^{\dagger}-\hat{a})].
\end{split}
\end{equation}
Neglect all the higher-order contributions, an effective Hamiltonian can be obtained~\cite{SupplementalMaterial}
\begin{equation}\label{eq:eq6}
\begin{split}
\hat{\mathcal{H}}_{\text{s}}^{\text{GRWA}}=&\tilde{\epsilon}\hat{J}_{z}+\Delta\hat{J}_{x}^{2}+\omega \hat{a}^{\dagger}\hat{a}+\frac{1}{2}\tilde{g}(\hat{J}_{-}\hat{a}^{\dagger}+\hat{J}_{+}\hat{a})\\
&+\frac{1}{2}\epsilon \Big{[}\hat{J}_{-}\hat{a}^{\dagger}F_{1}(\hat{a}^{\dagger}\hat{a})+\hat{J}_{+}F_{1}(\hat{a}^{\dagger}\hat{a})\hat{a}\Big{]},
\end{split}
\end{equation}
where $\hat{J}_{\pm}\equiv \hat{J}_{x}\pm i\hat{J}_{y}$, $\tilde{\epsilon}=\epsilon F_{0}(\hat{a}^{\dagger}\hat{a})$ is the renormalized frequency, $\Delta=\omega\lambda^{2}-2g\lambda$ and $\tilde{g}=g-\lambda\omega$. Here, $F_{n}(m)\equiv \frac{m!}{(m+n)!}e^{-\frac{1}{2}\lambda^{2}}\lambda^{n}L_{m}^{n}(\lambda^{2})$ with $L_{m}^{n}(\lambda^{2})$ being the Laguerre polynomials. Note that $\hat{\mathcal{H}}_{\text{s}}^{\text{GRWA}}$ is a block-diagonal matrix in the product basis, one can diagonalize it by hand, and the corresponding SNR can be analytically obtained. Next, we use several examples to illustrate the advantages of the strong coupling on the low-temperature frequency estimation with a Lorentzian spectral density~\cite{SupplementalMaterial}.

\section{Quantum Rabi model case}
For the finite-$N$ cases, $\hat{\mathcal{H}}_{\text{s}}$ is the famous quantum Rabi model~\cite{PhysRev.49.324,PhysRev.51.652,Xie_2017}. In Fig.~\ref{fig:fig2} (a) and (b), we display $\delta \mathbb{S}\equiv\mathbb{S}-\mathbb{S}_{\text{weak}}$ as a function of the temperature with $N=1$. As long as $\delta \mathbb{S}>0$, one concludes the strong coupling improves the metrology performance. Sharply contrary to the exponentially-decay behaviour in the weak-coupling case, we find $\mathbb{S}_{N=1}$ remains as a constant when the temperature approaches to zero. This result can be well-explained as follows. When $(E_{e}-E_{g})/E_{g}\ll 1$ with $E_{e}$ being the first excited energy, one can neglect the contributions from the high-energy eigenstates and reexpressed the partition function as $\mathcal{Z}\simeq \exp(-\beta E_{g})$. This result leads to
\begin{equation}\label{eq:eq7}
\lim_{T\rightarrow 0}\mathbb{S}_{N=1}\propto\frac{4(\partial_{\epsilon}^{2}E_{g})^{2}}{1-4(\partial_{\epsilon}E_{g})^{2}},
\end{equation}
which is independent of the temperature. The above low-temperature asymptotic result is confirmed by both the GRWA approach and the exact diagonalization method as displayed in Fig.~\ref{fig:fig2} (a) and (b).

The quantum Rabi model with $N=1$ can be experimentally simulated by means of the superconducting circuit, which allows the realization of strong and ultrastrong couplings. Using the experimentally available parameters in Ref.~\cite{Yoshihara2017}, we find $\mathbb{S}_{N=1}/\mathbb{S}_{\text{weak}}\simeq 936$ with $\epsilon/2\pi=3.84~\text{GHz}$, $\omega/2\pi=5.588~\text{GHz}$, $g/2\pi=5.63~\text{GHz}$ and $T=45~\text{mK}$. One sees that $\mathbb{S}_{N=1}$ is larger by almost 3 orders of magnitude than $\mathbb{S}_{\text{weak}}$, which evidently demonstrates the enhancement effect of strong couplings.

For the cases of $N=2$ and $N=3$, we find
\begin{equation}\label{eq:eq8}
\lim_{T\rightarrow 0}\mathbb{S}_{N\geq 2}\propto-\frac{1}{T}\partial_{\epsilon}^{2}E_{g}.
\end{equation}
Note that $\partial_{\epsilon}^{2}E_{g}<0$~\cite{SupplementalMaterial}, this result suggests a $T^{-1}$ scaling behaviour in low-temperature regimes. Such a scaling relation is confirmed by both the analytical and the numerical methods as displayed in Fig.~\ref{fig:fig2} (c)-(f). This surprised result means the metrological precision can be increased by decreasing the temperature in our strategy, which is quite similar to the Landau-bound-type scaling relation in the studies of temperature sensing~\cite{Paris_2016,PhysRevApplied.17.034073}. Moreover, we find that the $T^{-1}$-scaling relation seems universal for all the finite-$N$ cases~\cite{SupplementalMaterial}. This universality greatly expands the general applicability of our proposed scheme. All these results prove the error-divergence problem in the noisy frequency metrology can be completely eliminated by strong couplings. Next, we generalize our analysis to the limit of $N\rightarrow\infty$.

\begin{figure}
\centering
\includegraphics[angle=0,width=0.485\textwidth]{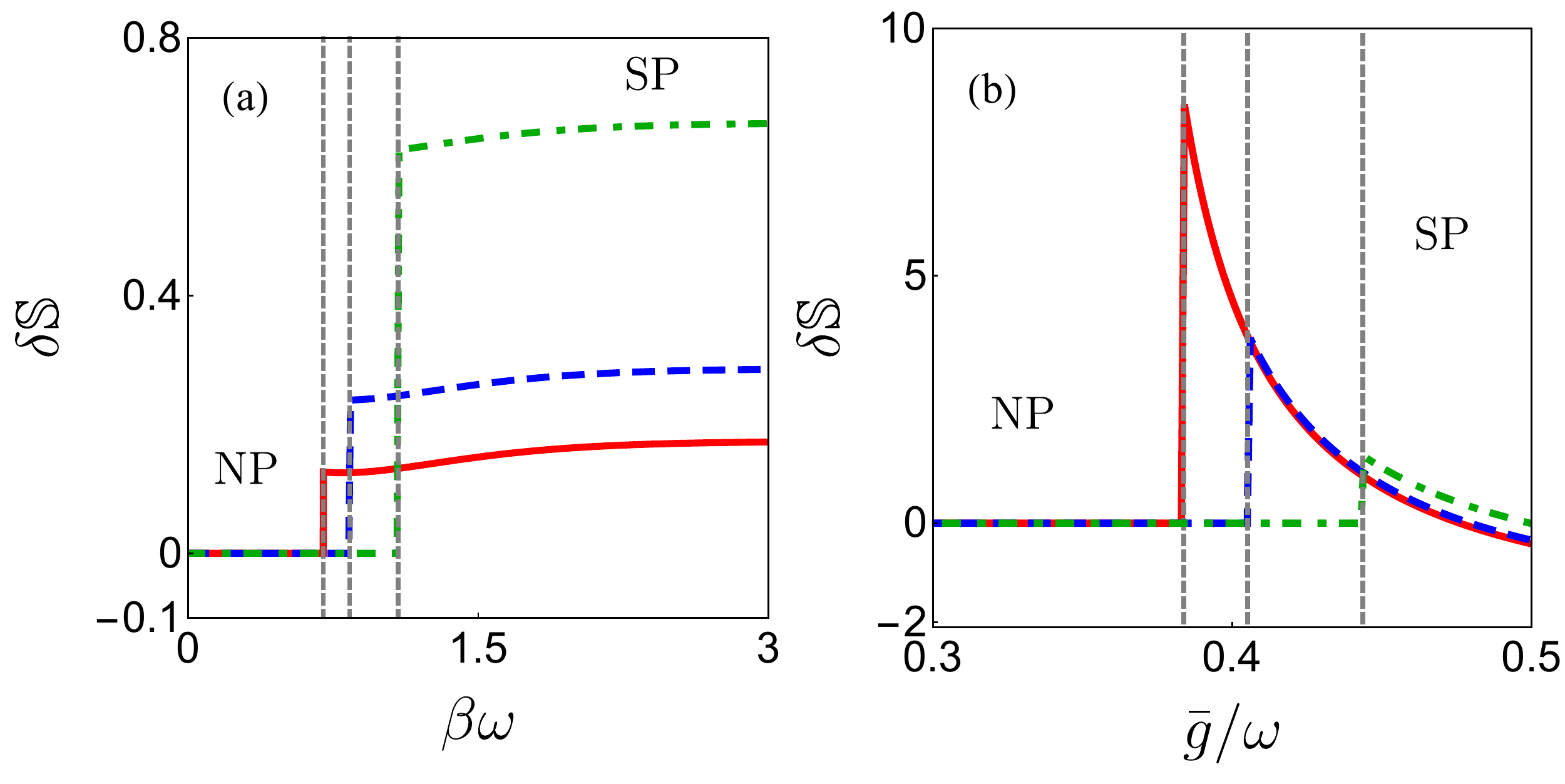}
\caption{(a) The modified SNR $\delta \mathbb{S}\equiv (\mathbb{S}_{\text{DM}}-\mathbb{S}_{\text{weak}})/N$ in the Dicke model case versus $\beta\omega$ with $\epsilon=3\omega$ for different coupling strengths: $\bar{g}/\omega=0.98$ (red soild line), $\bar{g}/\omega=0.94$ (blue dashed line) and $\bar{g}/\omega=0.9$ (green dot-dashed line). (b) $\delta \mathbb{S}$ versus $\bar{g}/\omega$ with $\epsilon=0.5\omega$ for different $\beta\omega$: $\beta\omega=5$ (red soild line), $\beta\omega=4$ (blue dashed line) and $\beta\omega=3$ (green dot-dashed line). The gray dotted lines separate the boundaries of the normal phase and the superradiant phase.}\label{fig:fig3}
\end{figure}

\section{Dicke model case}
In the limit of $N\rightarrow\infty$, $\hat{\mathcal{H}}_{\text{s}}$ becomes the famous quantum Dicke model~\cite{PhysRev.93.99}, whose thermodynamic properties has been widely discussed in previous studies~\cite{PhysRevA.7.831,PhysRevA.9.418,PhysRevA.70.033808,Liberti2005}. In this large-$N$ case, the excitation spectrum of the Dicke model becomes quasicontinuous, namely, the energy gap between the ground state and the first excited state becomes infinitesimal (almost gapless), which breaks down the condition $(E_{e}-E_{g})/E_{g}\ll 1$. Thus, neither the conclusions nor the methodology used in the finite-$N$ cases is directly applicable to the present situation. Fortunately, as displayed in Refs.~\cite{PhysRevA.7.831,PhysRevA.9.418,PhysRevA.70.033808,Liberti2005}, the Dicke model is exactly solvable in the limit $N\rightarrow\infty$. Via computing its partition function, the corresponding SNR can be analytically derived.

Following Refs.~\cite{PhysRevA.7.831,PhysRevA.9.418,PhysRevA.70.033808,Liberti2005}, the partition function of the Dicke model is given by~\cite{SupplementalMaterial}
\begin{equation}\label{eq:eq9}
\mathcal{Z}_{\text{DM}}=\sqrt{\frac{2}{\beta\omega|\partial^{2}_{z}\Phi(z)|}}e^{N\Phi(z)}\bigg{|}_{z=z_{0}},
\end{equation}
where $\Phi(z)=-\beta z^{2}+\ln[2\cosh(\frac{1}{2}\beta\sqrt{\epsilon^{2}+16\bar{g}^{2}z^{2}})]$ with $\bar{g}=\sqrt{N}g/2$ and $z_{0}$ being determined by the equation $\partial_{z}\Phi(z)|_{z=z_{0}}=0$. There are two possible roots for $\partial_{z}\Phi(z)|_{z=z_{0}}=0$, depending on the critical temperature
\begin{equation}\label{eq:eq10}
T_{\text{c}}^{\text{DM}}=\epsilon\bigg{[}2\mathrm{arctanh}\bigg{(}\frac{\epsilon\omega}{4\bar{g}^{2}}\bigg{)}\bigg{]}^{-1}
\end{equation}
at which the Dicke model experiences a thermodynamic phase transition. When $T>T_{\text{c}}^{\text{DM}}$, the Dicke model is in the normal phase with a trivial solution $z_{0}=0$ corresponding to the case in which the spins and the RC mode are completely decoupled. On the other hand, if $T\leq T_{\text{c}}^{\text{DM}}$, the Dicke model is in the superradiant phase with a nontrivial solution $z_{0}=\sqrt{\epsilon^{2}\eta^{2}-\epsilon^{2}}/(4\bar{g})$ where $\eta$ is determined by $\frac{1}{4}\eta\epsilon\omega\bar{g}^{-2}=\tanh(\frac{1}{2}\beta\eta\epsilon)$~\cite{PhysRevA.7.831,PhysRevA.9.418}.

With the above thermodynamic properties at hand, we find the SNR in the Dicke mode case is given by~\cite{SupplementalMaterial}
\begin{equation}\label{eq:eq11}
\frac{\mathbb{S}_{\text{DM}}}{N}=\begin{cases}
\beta^{2}/[2+2\cosh(\beta\epsilon)],&T>T_{\text{c}}^{\text{DM}};\\
\omega^{2}/(16\bar{g}^{4}-\epsilon^{2}\omega^{2}),&T\leq T_{\text{c}}^{\text{DM}}.\\
\end{cases}
\end{equation}
From the above expression, one sees $\mathbb{S}_{\text{DM}}$ in the normal phase has the same expression with that of the weak-coupling case, which means $\mathbb{S}_{\text{DM}}$ still suffers from the error-divergence problem at low temperatures. However, in the superradiant phase, the SNR becomes independent of the temperature circumventing the error-divergence problem. By engineering the parameters $\{\epsilon, \omega,\bar{g}\}$, the value of the SNR in the superradiant phase can be larger than $\mathbb{S}_{\text{weak}}$, as plotted in Fig.~\ref{fig:fig3}. Though the result of $\mathbb{S}_{\text{DM}}\propto T^{0}$ in the superradiant phase is similar to the case of Rabi model with $N=1$, these results are generated by different physical mechanisms. Moreover, when crossing over the phase boundary, one sees $\mathbb{S}_{\text{DM}}/N$ exhibits a discontinuous behavior resulting in a local maximum SNR at the phase transition point. Such a singularity is quite similar to previous studies of quantum critical metrology at zero temperature~\cite{PhysRevA.80.012318,PhysRevA.78.042106,PhysRevA.78.042105,Wang_2014,PhysRevE.93.052118,PhysRevX.8.021022,PhysRevLett.121.020402} and can be used to reveal the thermal phase transition without a prior knowledge about the order parameter or the symmetry. Our result suggests a phase transition, even happens at finite temperature, can be used as a resource to increase the metrology performance, which provides a possibility of realizing a quantum critical metrology without cooling down to $T\simeq0~\text{K}$.

\section{Conclusion}
In summary, we show that the error-divergence problem in a noisy frequency estimation task at low-temperature stems from the unnecessary weak-coupling approximation, which leads to a canonical thermalization for the probe within the Born-Markovian treatment. By employing the reaction-coordinate mapping, we overcome the restrict of the weak-coupling approximation and are able to study the influences of strong coupling, which naturally generates a noncanonical equilibrium state for the probe, on the metrology performance. By considering the strong-coupling effect, it is found that the SNR displays as a $T^{\theta}$-type scaling relation with $\theta=0$ for $N=1,~\infty$; and $\theta=-1$ for other finite-$N$ cases. This result is sharply contrary to the exponentially-decay SNR in the weak-coupling case. In this sense, we completely remove the error-divergence problem by the strong couplings. Paving a way to realize a high-precision quantum metrology at low temperature, our result reveals the importance of a proper understanding of equilibrium states in quantum technologies.

\section{Acknowledgments}
This work is supported by the National Natural Science Foundation of China (Grants No. 12375015, and No. 12247101).

\bibliography{reference}

\clearpage
\onecolumngrid
\begin{center}

{\large \bf Supplemental Materials for ``Low-temperature Quantum Metrology Enhanced by Strong Couplings'' }\\

\vspace{0.3cm}

\end{center}

This Supplemental Materials provides more details on the reaction-coordinate-mapping approach, the calculations of the SNR, the GRWA approach, the proof for the universality of the $T^{-1}$-scaling relation in the finite-N cases, as well as the thermodynamic properties of the Dicke model case.

\begin{center}

{\large \bf The reaction-coordinate-mapping approach}\\

\vspace{0.3cm}

\end{center}
The reaction-coordinate mapping can be viewed as a unitary transformation, which maps the original Hamiltonian
\begin{equation}
\hat{H}=\hat{H}_{\text{s}}+\sum_{k}\omega_{k}\hat{b}_{k}^{\dagger}\hat{b}_{k}+\hat{S}\sum_{k}g_{k}(\hat{b}^{\dagger}_{k}+\hat{b}_{k})
\end{equation}
into a new (mapped) Hamiltonian
\begin{equation}
\hat{\mathcal{H}}=\hat{H}_{\text{s}}+\omega \hat{a}^{\dagger}\hat{a}+g\hat{S}(\hat{a}^{\dagger}+\hat{a})+\sum_{k}\tilde{\omega}_{k}\hat{a}^{\dagger}\hat{a}_{k}+(\hat{a}^{\dagger}+\hat{a})\sum_{k}\tilde{g}_{k}(\hat{a}^{\dagger}_{k}+\hat{a}_{k})+(\hat{a}^{\dagger}+\hat{a})^{2}\sum_{k}\frac{\tilde{g}_{k}^{2}}{\tilde{\omega}_{k}}.
\end{equation}
In the main text, $H_{\text{s}}=\epsilon \hat{J}_{z}$ and $\hat{S}=\hat{J}_{x}$, but we want to emphasize that the reaction-coordinate-mapping approach is universal to different forms of $H_{\text{s}}$ and $\hat{S}$.

The relation between the spectral density of the original Hamiltonian $J(\varpi)=J^{(0)}(\varpi)=\sum_{k}g_{k}^{2}\delta(\varpi-\omega_{k})$ and the spectral density for the residual bath in $\mathcal{\hat{H}}$, which is defined by $\mathcal{J}(\varpi)=J^{(1)}(\varpi)=\sum_{k}\tilde{g}_{k}^{2}\delta(\varpi-\tilde{\omega}_{k})$, can be established via the dynamically equivalence between $\hat{H}$ and $\mathcal{\hat{H}}$~\cite{Nazir2018,PhysRevA.90.032114}. To see this, we first consider the Heisenberg equation of motion for an arbitrary system's operator $\hat{\mathcal{Q}}$ and the annihilate operators of the bath, namely $\hat{b}_{k}$, from the original Hamiltonian $\hat{H}$ as follows
\begin{align}
  &\dot{\hat{\mathcal{Q}}}=i[\hat{H},\hat{\mathcal{Q}}]=i\hat{\mathcal{Q}}_{1}+i\hat{\mathcal{Q}}_{0}\sum_{k}g_{k}(\hat{b}^{\dagger}_{k}+\hat{b}_{k}),\\
  &i\dot{\hat{b}}_{k}=i[\hat{H},\hat{b}_{k}]=-i\omega_{k}\hat{b}_{k}-ig_{k}\hat{S},
\end{align}
where $\hat{\mathcal{Q}}_{0}=[\hat{S},\hat{\mathcal{Q}}]$ and $\hat{\mathcal{Q}}_{1}=[\hat{H}_{\text{s}},\hat{\mathcal{Q}}]$. Applying the Laplace-Fourier transformation, which is introduced as $f(z)=\int_{0}^{\infty} f(t)e^{izt}dt$, to the above two equations, one sees
\begin{align}
  &iz\hat{\mathcal{Q}}(z)=i\hat{\mathcal{Q}}_{1}(z)+\frac{i}{2\pi}\int_{0}^{\infty} \hat{\mathcal{Q}}_{0}(z')\sum_{k}g_{k}\left[\hat{b}_{k}^{\dagger}(z-z')+\hat{b}_{k}(z-z')\right]dz',\\
  &iz\hat{b}_{k}(z)=-i\omega_{k}\hat{b}_{k}(z)-ig_{k}\hat{S}(z),
\end{align}
By solving the last equation with $\hat{b}_{k}(z)=-\frac{g_{k}}{z+\omega_{k}}\hat{S}(z)$, and inserting it into the first equation, one finds
\begin{equation}{\label{S1}}
  \begin{split}
    z\hat{\mathcal{Q}}(z)&=\hat{\mathcal{Q}}_{1}(z)+\frac{1}{2\pi}\int_{0}^{\infty} \hat{\mathcal{Q}}_{0}(z')\sum_{k}\left[ \frac{g_{k}^{2}}{z-z'-\omega_{k}}-\frac{g_{k}^{2}}{z-z'+\omega_{k}}\right]\hat{S}(z-z')dz',\\
    &=\hat{\mathcal{Q}}_{1}(z)+\frac{1}{2\pi}\int_{0}^{\infty}\hat{\mathcal{Q}}_{0}(z')\left[\frac{1}{\pi}\int_{0}^{\infty}J^{(0)}(\varpi)\frac{\varpi}{(z-z')^{2}-\varpi^{2}}d\varpi\right]\hat{S}(z-z')dz',\\
    &=\hat{\mathcal{Q}}_{1}(z)-\frac{1}{2\pi}\int_{0}^{\infty} \hat{\mathcal{Q}}_{0}(z')\frac{1}{2}W^{(0)}(z-z')\hat{S}(z-z')dz'.
  \end{split}
\end{equation}
Here, we have introduced the Cauchy transformation
\begin{equation}
  W^{(n)}(z)=\frac{2}{\pi}\int_{0}^{\infty}J^{(n)}(\varpi)\frac{\varpi}{\varpi^{2}-z^{2}}d\varpi=\frac{1}{\pi}\int_{-\infty}^{\infty}\frac{J^{(n)}(\varpi)}{\varpi-z}d\varpi,
\end{equation}
where we have extended the integral boundary via the analytic continuation by introducing $J^{(n)}(\varpi)=-J^{(n)}(-\varpi)$ for $\varpi<0$. Using the residue theorem, the spectral density $J^{(n)}(\varpi)$ can be expressed in terms of $W^{(n)}(z)$ as~\cite{Nazir2018,PhysRevA.90.032114,PhysRevB.30.1208}
\begin{equation}\label{Sj}
  J^{(n)}(\varpi)=\lim_{\delta\to 0^{+}}\text{Im}[W^{(n)}(\varpi+i\delta)].
\end{equation}

Using the same method, from the Heisenberg equation of motion for the mapped Hamiltonian $\hat{\mathcal{H}}$, one can find the equation of motion for an arbitrary system's operator $\hat{\mathcal{Q}}$ in the Laplace-Fourier space as~\cite{Nazir2018}
\begin{equation}{\label{S2}}
  z\hat{\mathcal{Q}}(z)=\hat{\mathcal{Q}}_{1}(z)+\frac{1}{2\pi}\int_{0}^{\infty} \hat{\mathcal{Q}}_{0}(z')\frac{2g^{2}\omega}{(z-z')^{2}-\omega^{2}+\omega W^{(1)}(z-z')}\hat{S}(z-z')dz'.
\end{equation}
Comparing Eq.~(\ref{S1}) and Eq.~(\ref{S2}), one can conclude that $\hat{H}$ and $\hat{\mathcal{H}}$ are dynamically equivalent if
\begin{equation}\label{S11}
 -\frac{1}{2}W^{(0)}(z-z')=\frac{2g^{2}\omega}{(z-z')^{2}-\omega^{2}+\omega W^{(1)}(z-z')}.
\end{equation}
By employing the above equation as well as Eq.~(\ref{Sj}), the relation between $J^{(0)}(\varpi)$ and $J^{(1)}(\varpi)$ can be built.

\begin{center}

    {\large \bf The spectral density}\\
    
    \vspace{0.3cm}
    
\end{center}

In our work, we assume the spectral density with respect to the mapped Hamiltonian $\hat{\mathcal{H}}$, which defined as $J^{(1)}(\varpi)=\sum_{k}\tilde{g}_{k}^{2}\delta(\varpi-\tilde{\omega}_{k})$, has an Ohmic form as
\begin{equation}
J^{(1)}(\varpi)=\gamma \varpi e^{-\varpi/\omega_{\text{c}}},
\end{equation}
where $\gamma$ is effective coupling strength and $\omega_{\text{c}}$ is cutoff frequency. With the help of the Eq.~(\ref{S11}), we find $J^{(0)}(\varpi)$ has a standard Lorentzian form as follows
\begin{equation}
\begin{split}
J^{(0)}(\varpi)=&\lim_{\delta\to 0^{+}}\text{Im}\frac{-4g^{2}\omega}{(\varpi+i\delta)^{2}-\omega^{2}+\omega W^{(1)}(\varpi+i\delta)}\\
=&\frac{4\gamma\omega^{2}g^{2}\varpi}{(\varpi^{2}-\omega^{2})^{2}+(\gamma\omega\varpi)^{2}}=\frac{\Gamma\varsigma\varpi}{(\varpi^{2}-\omega^{2})^{2}+\Gamma^{2}\varpi^{2}},
\end{split}
\end{equation}
where $\omega$ is the resonant frequency, $\Gamma=\gamma \omega$ is the peak width and $\varsigma=4\omega g^{2}$ can be viewed as the effective probe-bath coupling strength.

By appropriately choosing the parameters of $\{\varsigma,\Gamma,\omega\}$, one can realize an arbitrarily strong probe-bath coupling in the original Hamiltonian $\hat{H}$ with an arbitrarily large $\varsigma$, while having arbitrarily small coupling between $\mathcal{\hat{H}}_{\text{s}}$ and $\mathcal{\hat{H}}_{\text{b}}$ with an arbitrarily small $\gamma$. The above operation can be done because the condition of $\gamma\rightarrow 0$ is achievable via choosing $\Gamma/\omega\rightarrow 0$, which does not imply any constraint on the coupling strength $\varsigma$~\cite{Cerisola_2024}. Thus, in the limit $\gamma\rightarrow 0$, while allowing $\varsigma$ to be arbitrarily large, one can regard the extended composite system (the probe plus the RC mode) experiences a canonical thermalization, which results in $\varrho_{\text{s}}(\infty)=e^{-\beta \hat{\mathcal{H}}_{\text{s}}}/\mathcal{Z}$. Thus, the reaction-coordinate mapping recasts the non-Markovian dynamics of the probe into the Markovian dynamics of an enlarged system including the probe and an effective mode of the environment.

\begin{figure*}
\centering
\includegraphics[angle=0,width=0.90\textwidth]{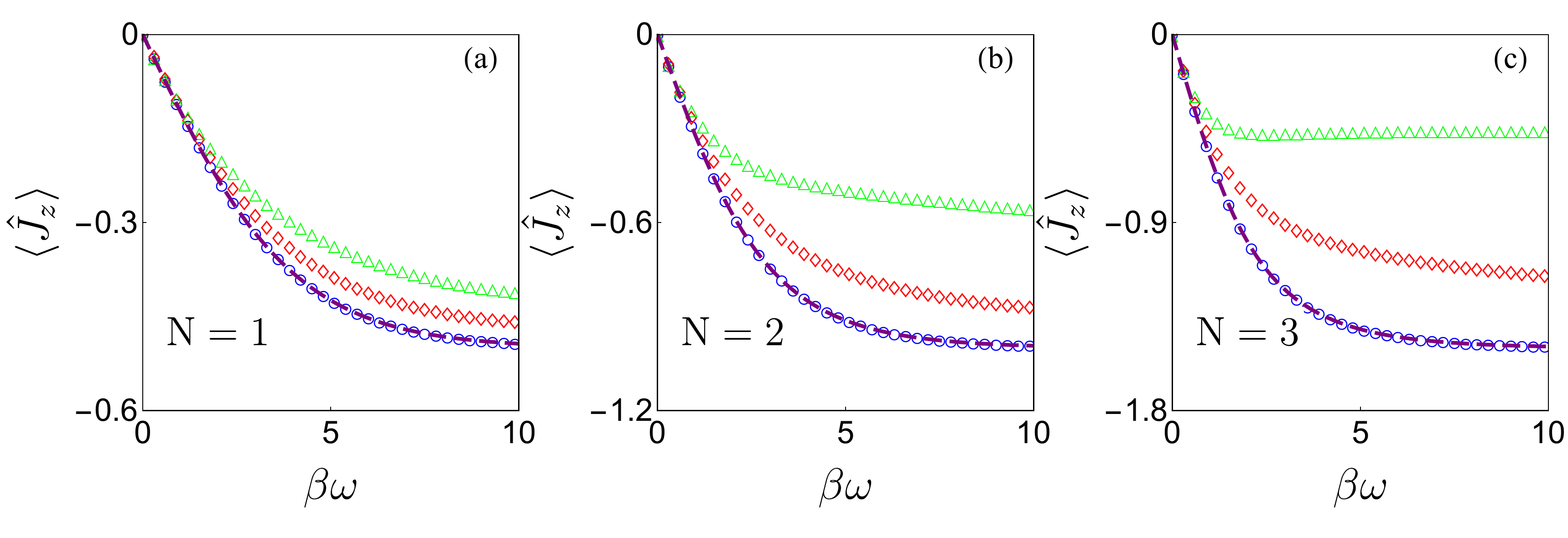}
\caption{The average values of $\langle \hat{J}_{z}\rangle$ are plotted as a function of the bath temperature $\beta\omega$ with $\epsilon=0.5\Delta$ for different spin numbers: (a) $N=1$, (b) $N=2$ and (c) $N=3$. The purple dashed lines are results from the weak-coupling treatment, while the blue circles ($g=0.01\omega$), the red diamonds ($g=0.5\omega$) and the green triangles ($g=0.8\omega$) are exact numerical results from the reaction coordinate mapping approach.}\label{fig:figs0}
\end{figure*}

In the especial limit $g\rightarrow 0$, which means $\varsigma\rightarrow 0$, one shall recover the usual weak-coupling result, i.e.,
\begin{equation}
\lim_{g\rightarrow 0}\text{Tr}_{\text{RC}}[\varrho_{\text{s}}(\infty)]=\rho_{\text{s}}^{\text{weak}}(\infty)=\frac{e^{-\beta \hat{H}_{\text{s}}}}{Z}.
\end{equation}
To check the above conclusion, in Fig.~\ref{fig:figs0}, we plot the average value of $\hat{J}_{z}$ from the weak-coupling approximation, i.e., $\langle \hat{J}_{z}\rangle_{\text{weak}}=\text{Tr}[\rho_{\text{s}}^{\text{weak}}(\infty)\hat{J}_{z}]$ and the reaction coordinate mapping approach, i.e., $\langle \hat{J}_{z}\rangle=\text{Tr}[\varrho_{\text{s}}(\infty)\hat{J}_{z}]$, versus the bath temperature. Good agreement is found between results from $\langle \hat{J}_{z}\rangle_{\text{weak}}$ and $\langle \hat{J}_{z}\rangle$ when $g$ is small. However, as the $g$ becomes large, a distinct deviation can be observed, which means the noncanonical effect induced by the strong couplings becomes non-negligible.

\begin{center}

    {\large \bf The SNR}\\
    
    \vspace{0.3cm}
    
\end{center}

By using the reaction-coordinate mapping, the long-time steady state of the probe in the strong-coupling regime is corrected as the reduced Gibbs state with respect to the Hamiltonian of the extended composite system, consisting of the probe and the RC mode, as~\cite{PhysRevA.90.032114,10.1063/1.4940218,10.1063/5.0207028,PhysRevE.110.014144,PhysRevA.108.032220}
\begin{equation}
\rho_{\text{s}}(\infty)=\text{Tr}_{\text{RC}}\bigg{(}\frac{e^{-\beta \hat{\mathcal{H}}_{\text{s}}}}{\mathcal{Z}}\bigg{)},
\end{equation}
where $\mathcal{Z}$ denotes the partition function $\mathcal{Z}=\text{Tr}(e^{-\beta \hat{\mathcal{H}}_{\text{s}}}$). Then, the expression of $\langle \hat{J}_{z}\rangle$ can be derived as
\begin{equation}
\langle \hat{J}_{z}\rangle=\text{Tr}_{\text{s}}[\rho_{\text{s}}(\infty)\hat{J}_{z}]=\frac{1}{\mathcal{Z}}\text{Tr}(e^{-\beta \hat{\mathcal{H}}_{\text{s}}}\hat{J}_{z})=-\frac{1}{\beta}\frac{\partial}{\partial\epsilon}\ln\mathcal{Z}.
\end{equation}
Similarly, one sees
\begin{equation}
\langle \hat{J}_{z}^{2}\rangle=\frac{1}{\mathcal{Z}\beta^{2}}\frac{\partial^{2}\mathcal{Z}}{\partial\epsilon^{2}}.
\end{equation}
Thus, as long as the partition function is obtained, the corresponding SNR can be accordingly derived. Technically speaking, the partition function can be obtained by numerically diagonalizing $\hat{\mathcal{H}}_{\text{s}}$. However, such a purely numerical treatment may miss some important physics. To build a more clear picture, in this section, we provide an analytical way to compute the partition function using the generalized rotating-wave approximation (GRWA) approach~\cite{PhysRevLett.99.173601,PhysRevA.91.013814,PhysRevA.94.012317}.

\begin{center}

    {\large \bf The Generalized Rotating-Wave Approximation}\\
    
    \vspace{0.3cm}
    
\end{center}

\begin{figure*}
\centering
\includegraphics[angle=0,width=0.90\textwidth]{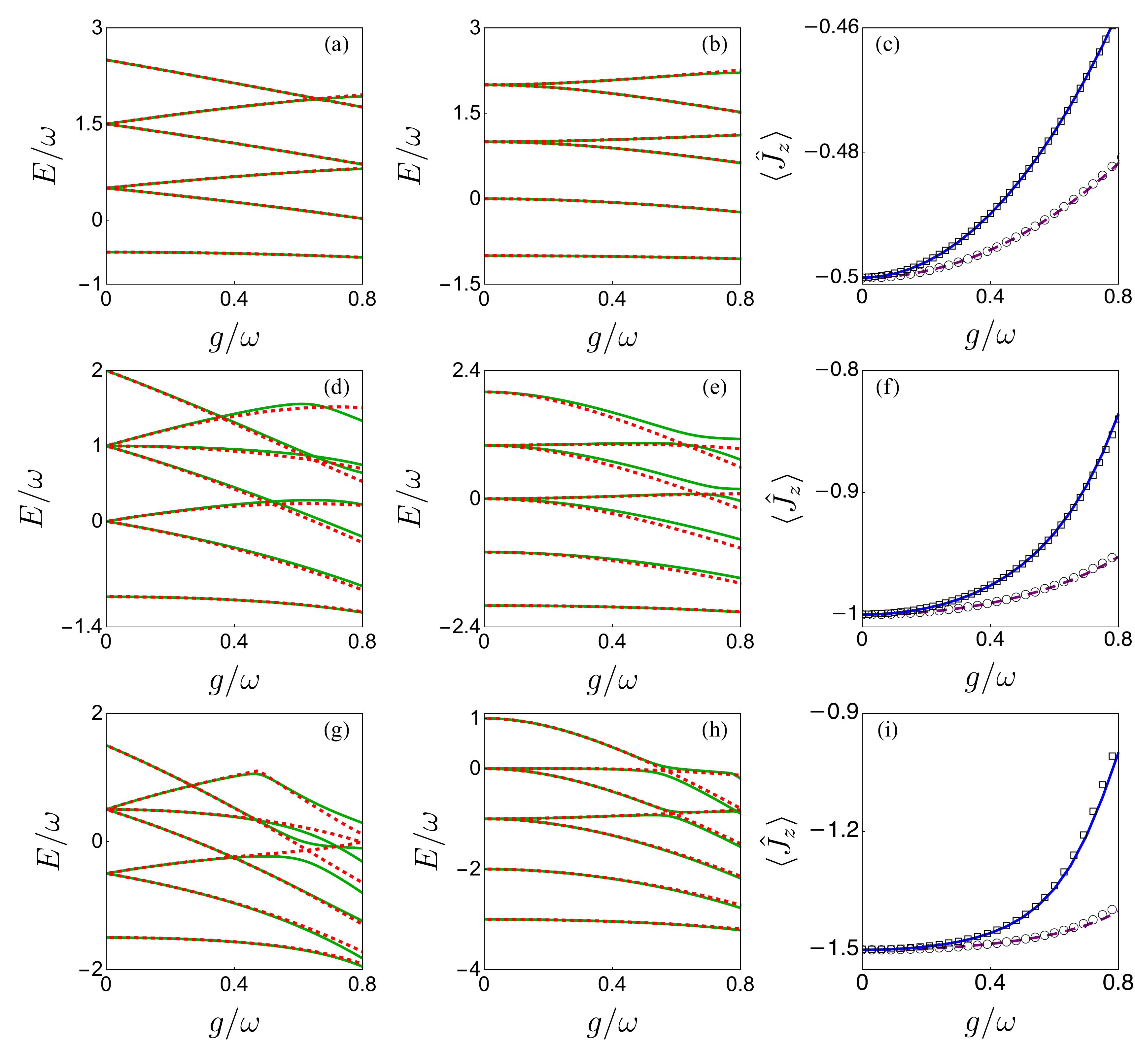}
\caption{Energy levels obtained by the GRWA approach (red dashed lines) and the numerically exact diagonalization (green solid lines) are plotted for comparison for different $\epsilon/\omega$ and $N=1$: (a) $\epsilon/\omega=1$ and $N=1$, (b) $\epsilon/\omega=2$ and $N=1$, (d) $\epsilon/\omega=1$ and $N=2$, (e) $\epsilon/\omega=2$ and $N=2$, (g) $\epsilon/\omega=1$ and $N=3$ and (h) $\epsilon/\omega=2$ and $N=3$. The average value of the observable $\langle \hat{J}_{z}\rangle$ obtained by the numerically exact diagonalization is plotted as a function of $g/\omega$ with different $\epsilon/\omega$: $\epsilon/\omega=1$ (blue solid lines) and $\epsilon/\omega=2$ (purple dashed lines). The squares and the circles are analytical results predicted by the GRWA approach.}\label{fig:figs1}
\end{figure*}

When the probe-RC mode coupling is not too strong, the character of the energy spectrum of $\hat{\mathcal{H}}_{\text{s}}$ can be accurately described by the GRWA approach (see the comparisons between the GRWA result and the exact result by numerically diagonalizing $\hat{\mathcal{H}}_{\text{s}}$ in Fig.~\ref{fig:figs1}.), which provides an analytical result maintaining strong links to the familiar language and techniques of quantum optics~\cite{PhysRevLett.99.173601}. To perform the GRWA approach, we first apply an unitary transformation to the Hamiltonian of $\hat{\mathcal{H}}_{\text{s}}$ as
\begin{equation}
\hat{\mathcal{H}}'_{\text{s}}=e^{\lambda \hat{J}_{x}(\hat{a}^{\dagger}-\hat{a})}\hat{\mathcal{H}}_{\text{s}}e^{-\lambda \hat{J}_{x}(\hat{a}^{\dagger}-\hat{a})},
\end{equation}
where $\lambda$ is a variational parameter and will be determined later. The transformed Hamiltonian $\hat{\mathcal{H}}'_{\text{s}}$ is given by
\begin{equation}
\hat{\mathcal{H}}'_{\text{s}}=\omega\hat{a}^{\dagger}\hat{a}+(\omega\lambda^{2}-2\lambda g)\hat{J}_{x}^{2}+(g-\omega\lambda)\hat{J}_{x}(\hat{a}^{\dagger}+\hat{a})+\epsilon\{ \hat{J}_{z}\cosh[\lambda(\hat{a}^{\dagger}-\hat{a})]-i\hat{J}_{y}\sinh[\lambda(\hat{a}^{\dagger}-\hat{a})]\}.
\end{equation}
Following Refs.~\cite{Liu_2015,PhysRevA.99.033834}, we expand the hyperbolic cosine and sine terms as follows
\begin{equation}
\cosh[\lambda(\hat{a}^{\dagger}-\hat{a})]=F_{0}(\hat{a}^{\dagger}\hat{a})+\sum_{n=1}^{\infty}[(a^{\dagger})^{2n}F_{2n}(\hat{a}^{\dagger}\hat{a})-F_{2n}(\hat{a}^{\dagger}\hat{a})\hat{a}^{2n}],
\end{equation}
\begin{equation}
\sinh[\lambda(\hat{a}^{\dagger}-\hat{a})]=\sum_{n=0}^{\infty}[(a^{\dagger})^{2n+1}F_{2n+1}(\hat{a}^{\dagger}\hat{a})-F_{2n+1}(\hat{a}^{\dagger}\hat{a})\hat{a}^{2n+1}],
\end{equation}
where the function $F_{n}(m)$ is defined by
\begin{equation}
F_{n}(m)\equiv \lambda^{n}e^{-\frac{1}{2}\lambda^{2}}\frac{m!}{(m+n)!}L_{m}^{n}(\lambda^{2})
\end{equation}
with $L_{m}^{n}(x)$ being the associated Laguerre polynomials
\begin{equation}
L_{m}^{n}(x)\equiv\sum_{j=0}^{m}\frac{(n+m)!}{(n+j)!(m-j)!j!}(-x)^{j}.
\end{equation}
Using these expanded expressions, one can drop all the higher-order terms of $(\hat{a}^{\dagger})^{l}\hat{a}^{l'}$ with $l,l'\geq 2$, which results in $\hat{\mathcal{H}}'_{\text{s}}\simeq \hat{\mathcal{H}}'_{0}+\hat{\mathcal{H}}'_{\text{RWA}}+\hat{\mathcal{H}}'_{\text{CRW}}$, where
\begin{equation}
\hat{\mathcal{H}}'_{0}=\omega\hat{a}^{\dagger}\hat{a}+(\omega\lambda^{2}-2\lambda g)\hat{J}_{x}^{2}+\epsilon\hat{J}_{z}F_{0}(\hat{a}^{\dagger}\hat{a}),
\end{equation}
is the zero-order (also called the adiabatic) term, and
\begin{equation}
\hat{\mathcal{H}}'_{\text{RWA}}=\frac{1}{2}(g-\omega\lambda)(\hat{J}_{-}\hat{a}^{\dagger}+\hat{J}_{+}\hat{a})+\frac{1}{2}\epsilon \Big{[}\hat{J}_{-}\hat{a}^{\dagger}F_{1}(\hat{a}^{\dagger}\hat{a})+\hat{J}_{+}F_{1}(\hat{a}^{\dagger}\hat{a})\hat{a}\Big{]},
\end{equation}
is the generalized rotating-wave-approximation term with $\hat{J}_{\pm}\equiv\frac{1}{2}(\hat{J}_{z}\pm i\hat{J}_{y})$, which conserves the total excitation and
\begin{equation}
\hat{\mathcal{H}}'_{\text{CRW}}=\frac{1}{2}(g-\omega\lambda)(\hat{J}_{-}\hat{a}+\hat{J}_{+}\hat{a}^{\dagger})-\frac{1}{2}\epsilon \Big{[}\hat{J}_{-}F_{1}(\hat{a}^{\dagger}\hat{a})\hat{a}+\hat{J}_{+}\hat{a}^{\dagger}F_{1}(\hat{a}^{\dagger}\hat{a})\Big{]},
\end{equation}
is the term containing the generalized counter-rotating-wave terms. In the GRWA treatment, one neglects the contribution from $\hat{\mathcal{H}}'_{\text{CRW}}$ and finally obtain the effective GRWA Hamiltonian as $\hat{\mathcal{H}}_{\text{GRWA}}=\hat{\mathcal{H}}'_{0}+\hat{\mathcal{H}}'_{\text{RWA}}$, which finally recovers Eq.~(6) in the main text.

Taking the direct product basis as $|\text{m},\text{n}\rangle\equiv|\text{m}\rangle\otimes|\text{n}\rangle$ with $\hat{J}_{z}|\text{m}\rangle=\text{m}|\text{m}\rangle$ and $\hat{a}^{\dagger}\hat{a}|\text{n}\rangle=\text{n}|\text{n}\rangle$, the approximate ground state of composite system $\hat{\mathcal{H}}_{\text{s}}$ within the treatment of the GRWA approach is then given by
\begin{equation}
|E_{g}\rangle\simeq|E_{g}^{\text{GRWA}}\rangle=e^{-\lambda \hat{J}_{x}(\hat{a}^{\dagger}-\hat{a})}\bigg{|}-\frac{N}{2},0\bigg{\rangle},
\end{equation}
and the corresponding ground-state energy reads
\begin{equation}
\begin{split}
E_{g}\simeq E_{g}^{\text{GRWA}}=&\bigg{\langle}-\frac{N}{2},0\bigg{|}\hat{\mathcal{H}}'_{0}+\hat{\mathcal{H}}'_{\text{RWA}}\bigg{|}-\frac{N}{2},0\bigg{\rangle}\\
=&\frac{N}{4}(\omega\lambda^{2}-2g\lambda)-\frac{N}{2}\epsilon e^{-\frac{1}{2}\lambda^{2}}.
\end{split}
\end{equation}
The other way to compute the ground-state energy is straightforwardly diagonalize $\hat{\mathcal{H}}'_{0}$ in the product basis $\{|\text{m},\text{n}\rangle\}$. These two methods yields the same physical results.

Up to here, the only task left to be completed is determining the variational parameter $\lambda$. Following Refs.~\cite{Liu_2015,PhysRevA.99.033834}, the parameter $\lambda$ is determined by minimizing the ground-state energy $E_{g}^{\text{GRWA}}$ as $\partial_{\lambda}E_{g}^{\text{GRWA}}=0$, which yields
\begin{equation}\label{eq:eqs17}
\lambda-\frac{g}{\omega}-\frac{\epsilon\lambda}{\omega}e^{-\frac{1}{2}\lambda^{2}}=0.
\end{equation}
By numerically solving the above transcendental equation, the value of $\lambda$ is then determined, which fully completes the GRWA approach. Moreover, Eq.~(\ref{eq:eqs17}) has an approximate solution as~\cite{Liu_2015,PhysRevA.99.033834}
\begin{equation}\label{eq:eqs18}
\lambda=\frac{g}{\omega+\epsilon e^{-\frac{1}{2}\lambda_{0}}},
\end{equation}
where $\lambda_{0}\equiv g/(\epsilon+\omega)$. This approximate solution can be used as a benchmark to testify the validity of purely numerical simulations. As plotted in Fig.~\ref{fig:figs2} (a), a good agreement is found between the approximate solution predicted by Eq.~(\ref{eq:eqs17}) and the purely numerical simulation from solving Eq.~(\ref{eq:eqs18}).

Next, we apply the GRWA approach to the cases of $N=1,2,3$ and display the analytical expressions of the partition function in the three cases, from which the SNRs can be accordingly derived.

\subsection{$N=1$ case}

For the simplest case $N=1$, the Hamiltonian of $\mathcal{\hat{H}}_{\text{GRWA}}^{N=1}$ is a block-diagonal matrix in the basis of $\{|g,\text{n}+1\rangle,|e,\text{n}\rangle\}$ with $|g\rangle$ and $|e\rangle$ being the eigenstates of Pauli operator $\hat{\sigma}_{z}$ as
\begin{equation}
\mathcal{\hat{H}}_{\text{GRWA}}^{N=1}=E_{g,\text{GRWA}}^{N=1}\mathbf{1}\oplus\mathfrak{\hat{H}}_{\text{n}=0}^{N=1}\oplus\mathfrak{\hat{H}}_{\text{n}=1}^{N=1}\oplus\mathfrak{\hat{H}}_{\text{n}=2}^{N=1}...,
\end{equation}
where $E_{g,\text{GRWA}}^{N=1}=\langle g,0|\mathcal{\hat{H}}_{\text{GRWA}}^{N=1}|g,0\rangle$ is the ground-state energy, $\mathbf{1}$ is a $1\times 1$ identity matrix and
with $\mathfrak{\hat{H}}_{\text{n}}^{N=1}$ being $2\times 2$ matrices as
\begin{equation}
\mathfrak{\hat{H}}_{\text{n}}^{N=1}=\left(
                                         \begin{array}{cc}
                                           \xi_{\text{n},+}^{N=1} & R_{\text{n}}^{N=1} \\
                                           R_{\text{n}}^{N=1} & \xi_{\text{n}+1,-}^{-} \\
                                         \end{array}
                                       \right),
\end{equation}
where
\begin{equation}
\xi_{\text{n},\pm}^{N=1}=\omega\text{n}+\frac{1}{4}(\omega\lambda^{2}-2g\lambda)\pm\frac{1}{2}\epsilon F_{0}(\text{n}),~~~R_{\text{n}}^{N=1}=\frac{1}{2}\sqrt{n+1}\Big{[}g-\omega\lambda+\epsilon F_{1}(\text{n})\Big{]}.
\end{equation}
By diagonalizing these $2\times 2$ matrices $\mathfrak{\hat{H}}_{\text{n}}^{N=1}$ by hand, the corresponding eigenvalues are
\begin{equation}
E_{\text{n},\pm}^{N=1}=\frac{1}{2}\bigg{[}\xi_{\text{n}}^{+}+\xi_{\text{n}+1}^{-}\pm\sqrt{4R_{\text{n}}^{2}+(\xi_{\text{n}}^{+}-\xi_{\text{n}+1}^{-})^{2}}\bigg{]}.
\end{equation}
Then, the partition function within the GRWA approach $\mathcal{Z}_{\text{GRWA}}^{N=1}=\text{Tr}(e^{-\beta \mathcal{\hat{H}}_{\text{GRWA}}^{N=1}})$ can be written as the sum of all these eigenenergies
\begin{equation}
\mathcal{Z}_{\text{GRWA}}^{N=1}=e^{-\beta E_{g,\text{GRWA}}^{N=1}}+\sum_{\text{n}=0}^{\infty}\sum_{\nu=\pm}e^{-\beta E_{\text{n},\nu}^{N=1}}.
\end{equation}
In Fig.~\ref{fig:figs1} (c), we display the results of $\langle \hat{J}_{z}\rangle$ calculated by the numerically exact diagonalization method and the GRWA approach. No distinctly difference is found, which convinces us that the GRWA method truly captures the essential physics of the quantum Rabi model.

\subsection{$N=2$ case}

In the case of $N=2$, the Hamiltonian of $\mathcal{\hat{H}}_{\text{GRWA}}^{N=2}$ is a block-diagonal matrix as
\begin{equation}
\mathcal{\hat{H}}_{\text{GRWA}}^{N=2}=E_{g,\text{GRWA}}^{N=2}\mathbf{1}\oplus\mathfrak{\hat{H}}_{\text{n}=0}^{N=2}\oplus\mathfrak{\hat{H}}_{\text{n}=1}^{N=2}\oplus\mathfrak{\hat{H}}_{\text{n}=2}^{N=2}...,
\end{equation}
where $\mathfrak{\hat{H}}_{\text{n}\geq1}^{N=2}$ are $3\times 3$ matrices in the basis of $\{|-1,\text{n}+1\rangle,|0,\text{n}\rangle,|1,\text{n}-1\rangle\}$
\begin{equation}
\mathfrak{\hat{H}}_{\text{n}\geq 1}^{N=2}=\left(
                                         \begin{array}{ccc}
                                           \xi_{\text{n}+1,-}^{N=2} & R_{\text{n},0}^{N=2} & 0 \\
                                           R_{\text{n},0}^{N=2} & \xi_{\text{n},0}^{N=2} & R_{\text{n},1}^{N=2} \\
                                           0 & R_{\text{n},1}^{N=2} & \xi_{\text{n}-1,+}^{N=2}
                                         \end{array}
                                       \right)
\end{equation}
with
\begin{equation}
\xi_{\text{n},\pm}^{N=2}=\omega\text{n}+\frac{1}{2}(\omega\lambda^{2}-2g\lambda)\pm F_{0}(\text{n}),~~~\xi_{\text{n},0}^{N=2}=\omega\text{n}+\omega\lambda^{2}-2g\lambda,
\end{equation}
\begin{equation}
R_{\text{n},0}^{N=2}=\frac{\sqrt{2(\text{n}+1)}}{2}\Big{[}g-\lambda\omega+\epsilon F_{1}(\text{n})\Big{]},~~~R_{\text{n},1}^{N=2}=\frac{\sqrt{2\text{n}}}{2}\Big{[}g-\lambda\omega+\epsilon F_{1}(\text{n}-1)\Big{]}.
\end{equation}
And $\mathfrak{\hat{H}}_{\text{n}=0}^{N=2}$ is a $2\times 2$ matrix in the basis $\{|-1,1\rangle,|0,0\rangle\}$ as
\begin{equation}
\mathfrak{\hat{H}}_{\text{n}=0}^{N=2}=\left(
                                        \begin{array}{cc}
                                          \xi_{0,0}^{N=2} & R_{0,0}^{N=2} \\
                                          R_{0,0}^{N=2} & \xi_{0,-}^{N=2} \\
                                        \end{array}
                                      \right).
\end{equation}
By diagonalizing these $2\times 2$ and $3\times 3$ matrices, the analytical expression of the partition function can be derived similar to the case of $N=1$.

\subsection{$N=3$ case}

For the $N=3$ case, the Hamiltonian of $\mathcal{\hat{H}}_{\text{GRWA}}^{N=3}$ is a block-diagonal matrix as
\begin{equation}
\mathcal{\hat{H}}_{\text{GRWA}}^{N=3}=E_{g,\text{GRWA}}^{N=3}\mathbf{1}\oplus\mathfrak{\hat{H}}_{\text{n}=0}^{N=3}\oplus\mathfrak{\hat{H}}_{\text{n}=1}^{N=3}\oplus\mathfrak{\hat{H}}_{\text{n}=2}^{N=3}\oplus\mathfrak{\hat{H}}_{\text{n}=3}^{N=3}...,
\end{equation}
where $\mathfrak{\hat{H}}_{\text{n}\geq 2}^{N=3}$ are $4\times 4$ matrices in the basis of $\{|-\frac{3}{2},\text{n}+2\rangle,|-\frac{1}{2},\text{n}+1\rangle,|\frac{1}{2},\text{n}\rangle,|\frac{3}{2},\text{n}-1\rangle\}$
\begin{equation}
\mathfrak{\hat{H}}_{\text{n}\geq 2}^{N=3}=\left(
                                         \begin{array}{cccc}
                                           \xi_{\text{n}+2,-}^{N=3} & R_{\text{n},0}^{N=3} & 0 & 0 \\
                                           R_{\text{n},0}^{N=3} & \zeta_{\text{n}+1,-}^{N=3} & R_{\text{n},1}^{N=3} & 0 \\
                                           0 & R_{\text{n},1}^{N=2} & \zeta_{\text{n},+}^{N=3} & R_{\text{n},2}^{N=3} \\
                                           0 & 0 & R_{\text{n},2}^{N=3} & \xi_{\text{n}-1,+}^{N=3}
                                         \end{array}
                                       \right),
\end{equation}
with
\begin{equation}
\xi_{\text{n},\pm}^{N=3}=\omega\text{n}+\frac{3}{4}(\omega\lambda^{2}-2g\lambda)\pm \frac{3}{2}\epsilon F_{0}(\text{n}),~~~\zeta_{\text{n},\pm}^{N=3}=\omega\text{n}+\frac{7}{4}(\omega\lambda^{2}-2g\lambda)\pm \frac{1}{2}\epsilon F_{0}(\text{n}),
\end{equation}
\begin{equation}
R_{\text{n},0}^{N=3}=\frac{\sqrt{3(\text{n}+2)}}{2}\Big{[}g-\lambda\omega+\epsilon F_{1}(\text{n}+1)\Big{]},~~R_{\text{n},1}^{N=3}=\sqrt{\text{n}+1}\Big{[}g-\lambda\omega+\epsilon F_{1}(\text{n})\Big{]},~~R_{\text{n},2}^{N=3}=\frac{\sqrt{3\text{n}}}{2}\Big{[}g-\lambda\omega+\epsilon F_{1}(\text{n}-1)\Big{]}.
\end{equation}
And $\mathfrak{\hat{H}}_{\text{n}=1}^{N=3}$ is a $3\times 3$ matrix in the basis $\{|-\frac{3}{2},2\rangle,|-\frac{1}{2},1\rangle,|\frac{1}{2},0\rangle\}$ as
\begin{equation}
\mathfrak{\hat{H}}_{\text{n}=1}^{N=3}=\left(
                                        \begin{array}{ccc}
                                          \xi_{2,-}^{N=3} & R_{0,0}^{N=3} & 0 \\
                                          R_{0,0}^{N=3} & \zeta_{1,-}^{N=3} & R_{0,1}^{N=3} \\
                                          0 & R_{0,1}^{N=3} & \zeta_{0,+}^{N=3}
                                        \end{array}
                                      \right).
\end{equation}
And $\mathfrak{\hat{H}}_{\text{n}=0}^{N=3}$ is a $2\times 2$ matrix in the basis $\{|-\frac{3}{2},1\rangle,|-\frac{1}{2},0\rangle\}$ as
\begin{equation}
\mathfrak{\hat{H}}_{\text{n}=0}^{N=3}=\left(
                                        \begin{array}{ccc}
                                          \xi_{1,-}^{N=3} & R_{-1,0}^{N=3} \\
                                          R_{-1,0}^{N=3} & \zeta_{0,-}^{N=3} \\
                                        \end{array}
                                      \right).
\end{equation}
By diagonalizing these $2\times 2$, $3\times 3$ and $4\times 4$ matrices, the analytical expression of the partition function can be derived similar to the case of $N=1$.

\begin{center}

    {\large \bf The Dicke model case}\\
    
    \vspace{0.3cm}
    
\end{center}

In this section, we shall discuss the thermodynamic properties of the Dicke model and derive the analytical expression of the SNR in the large-$N$ limit. To this aim, we first introduce a renormalized coupling strength as $\bar{g}=\sqrt{N}g/2$ and reexpress the Hamiltonian of the Dicke model as
\begin{equation}
\begin{split}
\hat{\mathcal{H}}_{\text{s}}=&\epsilon \hat{J}_{z}+\omega \hat{a}^{\dagger}\hat{a}+\frac{2\bar{g}}{\sqrt{N}}\hat{J}_{x}(\hat{a}^{\dagger}+\hat{a})\\
=&\sum_{n=1}^{N}\bigg{[}\frac{\epsilon}{2}\hat{\sigma}^{z}_{n}+\omega \frac{\hat{a}^{\dagger}}{\sqrt{N}}\frac{\hat{a}}{\sqrt{N}}
    +\frac{\bar{g}}{\sqrt{N}}\hat{\sigma}^{x}_{n}(\hat{a}+\hat{a}^{\dagger})\bigg{]}.
\end{split}
\end{equation}
Then, the partition function can be calculated as follows
\begin{equation}\label{eq:eq22}
\begin{split}
\mathcal{Z}_{\text{DM}}=\text{Tr}\Big{(}e^{-\beta\hat{\mathcal{H}}_{\text{s}}}\Big{)}=\sum_{\sigma_{1}=g,e}\sum_{\sigma_{2}=g,e}...\sum_{\sigma_{N}=g,e}\langle\sigma_{1}\sigma_{2}...\sigma_{N}|\int_{-\infty}^{+\infty} \frac{d^{2}\alpha}{\pi}\langle\alpha|e^{-\beta\hat{\mathcal{H}}_{\text{s}}}|\alpha\rangle|\sigma_{1}\sigma_{2}...\sigma_{N}\rangle,
\end{split}
\end{equation}
where $|\alpha\rangle$ introduced as the bosonic coherent state $\hat{a}|\alpha\rangle=\alpha|\alpha\rangle$. In the large-$N$ limit, we have $\sqrt{N}\gg\max\{\epsilon,\omega,g\}$ which leads to
\begin{equation}\label{eq:eq23}
\begin{split}
\langle\alpha|e^{-\beta\hat{\mathcal{H}}_{\text{s}}}|\alpha\rangle&\simeq\prod_{n}\bigg{\langle}\alpha\bigg{|}\exp\bigg{\{}-\beta\bigg{[}\frac{\epsilon}{2}\hat{\sigma}^{z}_{n}+\omega \frac{\hat{a}^{\dagger}}{\sqrt{N}}\frac{\hat{a}}{\sqrt{N}}
    +\frac{\bar{g}}{\sqrt{N}}\hat{\sigma}^{x}_{n}(\hat{a}+\hat{a}^{\dagger})\bigg{]}\bigg{\}}\bigg{|}\alpha\bigg{\rangle}\\
    &\simeq\prod_{n}\exp\bigg{\{}-\beta\bigg{\langle}\alpha\bigg{|}\bigg{[}\frac{\epsilon}{2}\hat{\sigma}^{z}_{n}+\omega \frac{\hat{a}^{\dagger}}{\sqrt{N}}\frac{\hat{a}}{\sqrt{N}}
    +\frac{\bar{g}}{\sqrt{N}}\hat{\sigma}^{x}_{n}(\hat{a}+\hat{a}^{\dagger})\bigg{]}\bigg{|}\alpha\bigg{\rangle}\bigg{\}}\\
    &=e^{-\beta\omega|\alpha|^{2}}\prod_{n}e^{-\beta \hat{O}_{n}},
\end{split}
\end{equation}
where
\begin{equation}\label{eq:eq24}
\hat{O}_{n}=\frac{\epsilon}{2}\hat{\sigma}^{z}_{n}+\frac{2\bar{g}\text{Re}\alpha}{\sqrt{N}}\hat{\sigma}^{x}_{n}.
\end{equation}
Thus, we have
\begin{equation}\label{eq:eq25}
\begin{split}
\mathcal{Z}_{\text{DM}}\simeq&\int_{-\infty}^{+\infty} \frac{d^{2}\alpha}{\pi}e^{-\beta\omega|\alpha|^{2}}\bigg{(}\sum_{\sigma=\uparrow\downarrow}\langle\sigma|e^{-\beta\hat{O}_{n}}|\sigma\rangle\bigg{)}^{N}\\
=&\int_{-\infty}^{+\infty} \frac{d^{2}\alpha}{\pi}e^{-\beta\omega|\alpha|^{2}}\Bigg{\{}2\cosh\Bigg{[}\beta\sqrt{\frac{\epsilon^{2}}{4}+\frac{4\bar{g}^{2}(\mathrm{Re}\alpha)^{2}}{N}}\Bigg{]}\Bigg{\}}^{N}.
\end{split}
\end{equation}
To handle the complex integral, we introduce $x\equiv\mathrm{Re}\alpha$ and $y\equiv\mathrm{Im}\alpha$, which results in $d^{2}\alpha=dxdy$ and $|\alpha|^{2}=x^{2}+y^{2}$. By doing so, the $y$-part of the integral is a Gaussian integral and can be immediately carried out. Then, one finds
\begin{equation}
\begin{split}
\mathcal{Z}_{\mathrm{DM}}=&\frac{1}{\sqrt{\pi\beta\omega}}\int_{-\infty}^{\infty}dxe^{-\beta\omega x^{2}}\Bigg{[}2\cosh\Bigg{(}\beta\sqrt{\frac{\epsilon^{2}}{4}+\frac{4\bar{g}^{2}x^{2}}{N}}\Bigg{)}\Bigg{]}^{N}.
\end{split}
\end{equation}
The above expression is still intricate. We further use the steepest descent method or the so-called Laplace's integral method~\cite{PhysRevA.70.033808,Liberti2005,PhysRevE.106.034104} to derive an approximate expression. To this aim, we replace $x/\sqrt{N}$ by a new variable $z$, then the expression of $\mathcal{Z}_{\mathrm{DM}}$ can be rewritten as
\begin{equation}\label{eq:eqs39}
\begin{split}
\mathcal{Z}_{\mathrm{DM}}=&\sqrt{\frac{N}{\pi\beta\omega}}\int_{-\infty}^{\infty}dze^{N\Phi(z)},
\end{split}
\end{equation}
where $\Phi(z)=-\beta z^{2}+\ln[2\cosh(\frac{1}{2}\beta\sqrt{\epsilon^{2}+16\bar{g}^{2}z^{2}})]$. The form of the partition function in Eq.~(\ref{eq:eqs39}) is especially suitable for the Laplace's integral method, which consists in approximating the exponential integrand by a Gaussian function around the global maximum of the function $\Phi(z)$. By employing the Laplace approximation, one can finally obtain
\begin{equation}
\begin{split}
\mathcal{Z}_{\mathrm{DM}}\simeq\sqrt{\frac{2}{\beta\omega|\partial_{z}^{2}\Phi(z)|}}e^{N\Phi(z)}\bigg{|}_{z=z_{0}},
\end{split}
\end{equation}
where $z_{0}$ is determined by $\partial_{z}\Phi(z)|_{z=z_{0}}=0$.

With the analytical expression of the Dicke model at hand, the average value of the chosen observable $\hat{J}_{z}$ can be calculate as
\begin{equation}
\langle \hat{J}_{z}\rangle=-\frac{1}{\beta}\frac{\partial}{\partial\epsilon}\ln \mathcal{Z}_{\mathrm{DM}}=\begin{cases}
-N\tanh(\beta\epsilon/2)/2,&T>T_{\text{c}}^{\text{DM}};\\
-N\tanh(\beta\epsilon\eta/2)/(2\eta),&T\leq T_{\text{c}}^{\text{DM}}.\\
\end{cases}
\end{equation}
and
\begin{equation}
\langle \hat{J}_{z}^{2}\rangle=\frac{1}{\beta^{2}\mathcal{Z}_{\text{DM}}}\frac{\partial^{2}\mathcal{Z}_{\mathrm{DM}}}{\partial\epsilon^{2}}=\begin{cases}
N/4+N(N-1)\tanh(\beta\epsilon/2)/4,&T>T_{\text{c}}^{\text{DM}};\\
N/4+N(N-1)\tanh(\beta\epsilon\eta/2)/(4\eta^{2}),&T\leq T_{\text{c}}^{\text{DM}}.\\
\end{cases}
\end{equation}
These results recover Eq.~(11) in the main text.

\begin{center}

    {\large \bf The universality of the $T^{-1}$-type scaling relation}\\
    
    \vspace{0.3cm}
    
\end{center}

\begin{figure*}
\centering
\includegraphics[angle=0,width=0.70\textwidth]{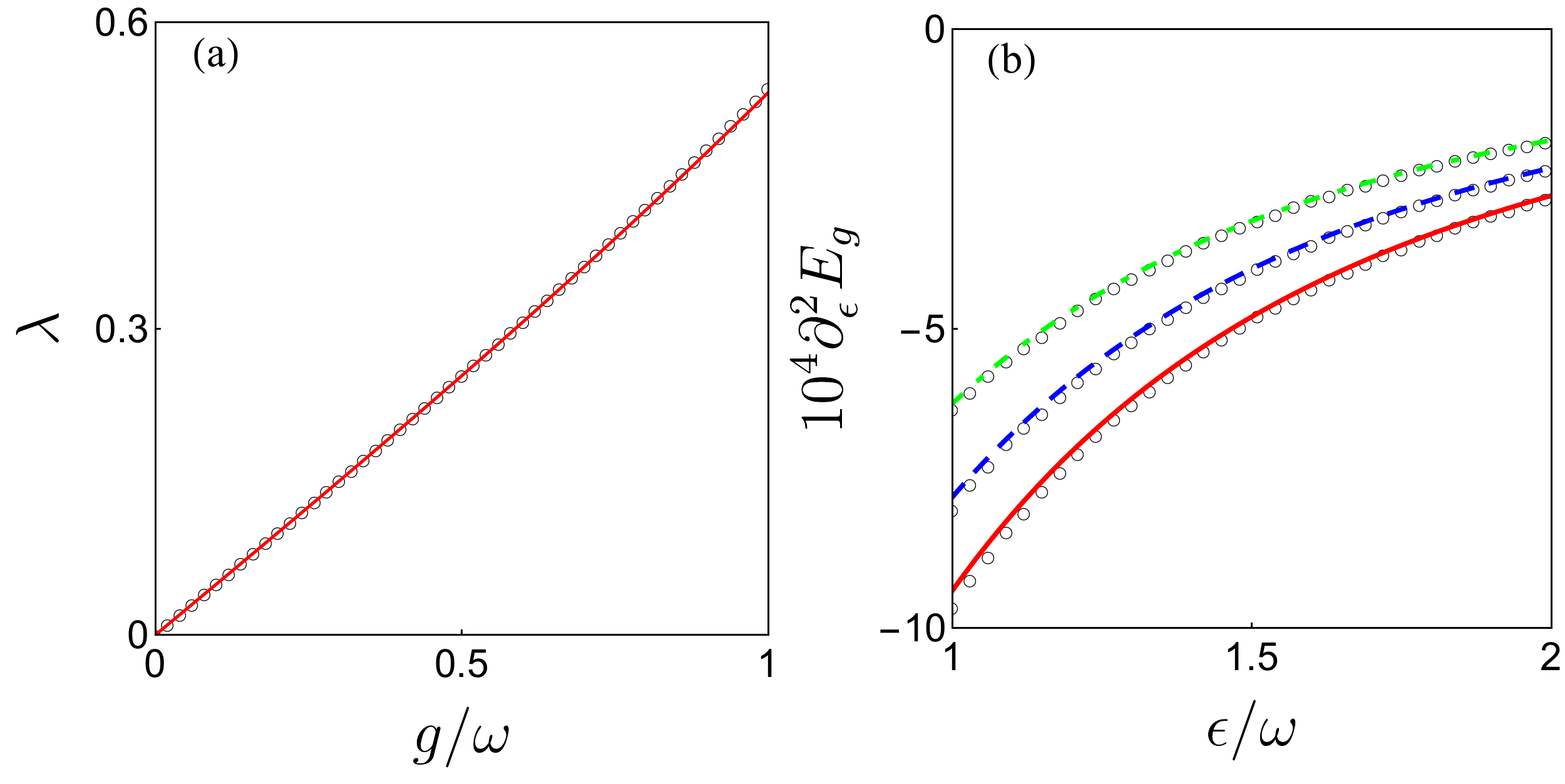}
\caption{(a) The variational parameter $\lambda$ is plotted as a function of $g/\omega$ with $\epsilon=\omega$. The circles are the result from numerically solving the above transcendental equation Eq.~(\ref{eq:eqs17}), while the red solid line is analytical prediction by Eq.~(\ref{eq:eqs18}). (b) $\partial^{2}_{\epsilon}E_{g}$ versus $\epsilon/\omega$ with $g=0.05\omega$ for different $N$: $N=4$ (green dotdashed line), $N=5$ (blue dashed line) and $N=6$ (red solid line). The circles are numerical results from the exact diagonalization, while the lines are analytical prediction from the GRWA. }\label{fig:figs2}
\end{figure*}

In this section, we should provide the proof that $\partial_{\epsilon}^{2}E_{g}<0$ within the framework of the GRWA approach for the finite-$N$ cases. Using the Feynman-Hellman theorem, one sees
\begin{equation}
\begin{split}
\frac{\partial^{2}}{\partial\epsilon^{2}}E_{g}^{\text{GRWA}}=&\frac{\partial}{\partial\epsilon}\langle E_{g}^{\text{GRWA}}|\hat{J}_{z}|E_{g}^{\text{GRWA}}\rangle\\
=&\frac{\partial}{\partial\epsilon}\bigg{\langle}-\frac{N}{2},0\bigg{|}e^{\lambda \hat{J}_{x}(\hat{a}^{\dagger}-\hat{a})}\hat{J}_{z}e^{-\lambda \hat{J}_{x}(\hat{a}^{\dagger}-\hat{a})}\bigg{|}-\frac{N}{2},0\bigg{\rangle}\\
=&\frac{N}{2}\lambda e^{-\frac{1}{2}\lambda^{2}}\frac{\partial\lambda}{\partial\epsilon}.
\end{split}
\end{equation}
Together with the approximate solution of the variational parameter $\lambda$ predicted by Eq.~(\ref{eq:eqs18}), one finds
\begin{equation}
\frac{\partial\lambda}{\partial\epsilon}=-\frac{g e^{\frac{1}{2}\lambda_{0}^{2}}[\epsilon g^{2}+(\epsilon+\omega)^{3}]}{(\epsilon+\omega)^{3}(\epsilon+\omega e^{\frac{1}{2}\lambda_{0}^{2}})^{2}}<0.
\end{equation}
Thus, we finally prove that $\partial_{\epsilon}^{2}E_{g}^{\text{GRWA}}<0$. Going beyond the GRWA treatment, we also provide the numerical evaluations of $\lambda$ by using exact diagonalization method with $N=4,5,6$ in Fig.~\ref{fig:figs2} (b). These numerical simulations confirm our analytical predictions from the GRWA approach. The proof of $\partial_{\epsilon}^{2}E_{g}<0$ validates the universality of $T^{-1}$-type scaling relation in the finite-$N$ cases, which plays a crucial role in our strategy to overcome the error-divergence problem at low temperature.

\section{The Excitation spectrum of the Dicke model}
In this section, we would like to show that the excitation spectrum of the Dicke model becomes quasicontinuous in the limit $N\rightarrow\infty$. To this aim, we use the Holstein-Primakoff transformation~\cite{PhysRev.58.1098} which represents the collective spin operators in terms of bosonic operators as follows~\cite{PhysRevE.67.066203}
\begin{equation}
\hat{J}_{+}=\hat{c}^{\dagger}\sqrt{N-\hat{c}^{\dagger}\hat{c}},~~~\hat{J}_{-}=\sqrt{N-\hat{c}^{\dagger}\hat{c}}\hat{c},~~~\hat{J}_{z}=\hat{c}^{\dagger}\hat{c}-\frac{N}{2},
\end{equation}
where $\hat{c}$ is introduced as a bosonic annihilate operator and obeys $[\hat{c},\hat{c}^{\dagger}]=1$. After the Holstein-Primakoff transformation, the Hamiltonian of the Dicke model becomes
\begin{equation}
\mathcal{\hat{H}}_{\text{s}}=\epsilon\Bigg{(}\hat{c}^{\dagger}\hat{c}-\frac{N}{2}\Bigg{)}+\omega \hat{a}^{\dagger}\hat{a}+\bar{g}(\hat{a}^{\dagger}+\hat{a})\Bigg{(}\hat{c}^{\dagger}\sqrt{1-\frac{\hat{c}^{\dagger}\hat{c}}{N}}+\sqrt{1-\frac{\hat{c}^{\dagger}\hat{c}}{N}}\hat{c}\Bigg{)}.
\end{equation}
In the normal phase, one has $\hat{c}^{\dagger}\hat{c}/N\rightarrow0$, which leads to
\begin{equation}
\mathcal{\hat{H}}_{\text{NP}}=\epsilon\hat{c}^{\dagger}\hat{c}+\omega \hat{a}^{\dagger}\hat{a}+\bar{g}(\hat{a}^{\dagger}+\hat{a})(\hat{c}^{\dagger}+\hat{c})-\frac{1}{2}N\epsilon,
\end{equation}
which is bilinear in terms of bosonic operators and can be diagonalized via the Bogoliubov transformation. After the Bogoliubov transformation, we have
\begin{equation}
\mathcal{\hat{H}}_{\text{NP}}=\sum_{\nu=\pm}\varepsilon_{\nu}^{\text{NP}}\bigg{(}\hat{d}_{\nu}^{\dagger}\hat{d}_{\nu}+\frac{1}{2}\bigg{)}-\frac{1}{2}(\epsilon+\omega)-\frac{1}{2}N\epsilon,
\end{equation}
where the excitation energies are
\begin{equation}
(\varepsilon_{\pm}^{\text{NP}})^{2}=\frac{1}{2}\Big{\{}\epsilon^{2}+\omega^{2}\pm\sqrt{(\epsilon^{2}-\omega^{2})^{2}+16\bar{g}^{2}\epsilon\omega}\Big{\}}.
\end{equation}

In the super-radiant phase, to incorporate the effect that both the spin ensemble and the RC mode acquire macroscopic occupations, one needs to displace the bosonic modes in the way as
\begin{equation}
\hat{a}\rightarrow \hat{e}-\frac{N}{\omega}\sqrt{\frac{N}{4}(1-\mu^{2})},~~~\hat{c}\rightarrow \hat{f}+\sqrt{\frac{N}{2}(1-\mu)},
\end{equation}
where $\mu\equiv\epsilon\omega/(4\bar{g}^{2})$. Then, the Holstein-Primakoff transformed Hamiltonian becomes
\begin{equation}
\begin{split}
\hat{\mathcal{H}}_{\text{SP}}=&\omega \hat{e}^{\dagger}\hat{e}+\frac{\epsilon(1+\mu)}{2\mu}\hat{f}^{\dagger}\hat{f}+\frac{\epsilon(1-\mu)(3+\mu)}{8\mu(1+\mu)}(\hat{f}+\hat{f}^{\dagger})^{2}\\
&+\bar{g}\mu\sqrt{\frac{2}{1+\mu}}(\hat{e}+\hat{e}^{\dagger})(\hat{f}+\hat{f}^{\dagger})-\frac{N}{2}\bigg{(}\frac{2\bar{g}^{2}}{\omega}+\frac{\epsilon^{2}\omega}{8\bar{g}^{2}}\bigg{)}-\frac{\bar{g}^{2}}{\omega}(1-\mu).
\end{split}
\end{equation}
Using the Bogoliubov transformation, the above expression can be rewritten in a diagonal form as
\begin{equation}
\hat{\mathcal{H}}_{\text{SP}}=\sum_{\nu=\pm}\varepsilon_{\nu}^{\text{SP}}\bigg{(}\hat{h}_{\nu}^{\dagger}\hat{h}_{\nu}+\frac{1}{2}\bigg{)}-\frac{1}{2}\bigg{[}\frac{\epsilon(1+\mu)}{2\mu}+\omega+\frac{2\bar{g}^{2}(1-\mu)}{\omega}\bigg{]}-\frac{N}{2}\bigg{(}\frac{2\bar{g}^{2}}{\omega}+\frac{\epsilon^{2}\omega}{8\bar{g}^{2}}\bigg{)},
\end{equation}
where the excitation energies in the super-radiant phase are given by
\begin{equation}
(\varepsilon_{\pm}^{\text{NP}})^{2}=\frac{1}{2}\frac{\epsilon^{2}}{\mu^{2}}+\omega^{2}\pm\sqrt{\bigg{(}\frac{\epsilon^{2}}{\mu^{2}}-\omega^{2}\bigg{)}^{2}+4\epsilon^{2}\omega^{2}}.
\end{equation}
From these expressions, one sees that, irrespective of whether the Dicke model is in the normal phase or in the super-radiant phase, the ground-state energy $E_{g}$ has the order of $O(N)$, while both the excitation energies $\varepsilon_{\nu}^{\text{NP}}$ and $\varepsilon_{\nu}^{\text{SP}}$ are $O(1)$, which leads to
\begin{equation}
\frac{E_{e}-E_{g}}{E_{g}}\simeq 1
\end{equation}
in the limit $N\rightarrow\infty$. This expression means the excitation spectrum of the Dicke model becomes quasicontinuous in the limit $N\rightarrow\infty$~\cite{PhysRevE.67.066203} and the condition of $(E_{e}-E_{g})/E_{g}\ll 1$, which is used in the analysis of the quantum Rabi model case, completely breaks down.

\end{document}